%% file: MOFHEI.tex
\documentclass[conference]{IEEEtran}
\usepackage{booktabs,caption}
\usepackage{graphicx}
\usepackage{caption}
\usepackage{subcaption}
\usepackage{amsmath}
\usepackage[flushleft]{threeparttable}
\usepackage{multirow}
\usepackage{algorithm}
\usepackage{algpseudocode}
\usepackage{url}
\usepackage{hyperref}
\usepackage{tablefootnote}
\ifCLASSINFOpdf
\else
\fi
\begin{document}
%
\title{MOFHEI: \underline{M}odel \underline{O}ptimizing \underline{F}ramework for Fast and Efficient \underline{H}omomorphically \underline{E}ncrypted Neural Network \underline{I}nference}



%
\author{\IEEEauthorblockN{Parsa Ghazvinian\IEEEauthorrefmark{1},
Robert Podschwadt\IEEEauthorrefmark{2},
Prajwal Panzade\IEEEauthorrefmark{1},
Mohammad H. Rafiei\IEEEauthorrefmark{3} and
Daniel~Takabi\IEEEauthorrefmark{2}}
\IEEEauthorblockA{\IEEEauthorrefmark{1}Georgia State University, 
Emails: pghazvinian1, ppanzade1 @gsu.edu}
\IEEEauthorblockA{\IEEEauthorrefmark{2}Old Dominion University, 
Emails: rpodschw, takabi @odu.edu}
\IEEEauthorblockA{\IEEEauthorrefmark{3}Johns Hopkins University,
Email: mrafiei1@jhu.edu}
}


\IEEEoverridecommandlockouts
\makeatletter\def\@IEEEpubidpullup{6.5\baselineskip}\makeatother

\maketitle

\begin{abstract}
Due to the extensive application of machine learning (ML) in a wide range of fields and the necessity of data privacy, privacy-preserving machine learning (PPML) solutions have recently gained significant traction. One group of approaches relies on Homomorphic Encryption (HE), which enables us to perform ML tasks over encrypted data. However, even with state-of-the-art HE schemes, HE operations are still significantly slower compared to their plaintext counterparts and require a considerable amount of memory. Therefore, we propose MOFHEI, a framework that optimizes the model to make HE-based neural network inference, referred to as private inference (PI), fast and efficient. First, our proposed learning-based method automatically transforms a pre-trained ML model into its compatible version with HE operations, called the HE-friendly version. Then, our iterative block pruning method prunes the model's parameters in configurable block shapes in alignment with the data packing method. This allows us to drop a significant number of costly HE operations, thereby reducing the latency and memory consumption while maintaining the model's performance. We evaluate our framework through extensive experiments on different models using various datasets. Our method achieves up to 98\% pruning ratio on LeNet, eliminating up to 93\% of the required HE operations for performing PI, reducing latency and the required memory by factors of 9.63 and 4.04, respectively, with negligible accuracy loss.
\end{abstract}


%

\section{Introduction}
\label{sec:Intro}

Due to the outstanding performance of machine learning (ML) solutions, they have been employed in a vast range of areas to leverage their capability in extracting complex patterns and insights from massive data \cite{schroff2015facenet}. However, training efficient ML models demands large amounts of datasets, considerable computation power, and ML expertise. To address these issues, cloud providers offer Machine Learning as a Service (MLaaS), which provides ingenious trained models and computation power, relieving clients from training complex models from scratch. Clients need to send their data to the cloud server to benefit from the offered utility \cite{wang2018rafiki}. Nevertheless, outsourcing sensitive data like medical or financial records introduces new data privacy concerns and does not comply with regulations such as GDPR \cite{gdpr} and HIPAA \cite{hipaa}. Privacy-Preserving Machine Learning (PPML) techniques have been proposed to address this privacy issue by employing techniques like differential privacy \cite{wang2023differential}, federated learning \cite{mcmahan2017communication}, Secure Multi-party Computation (SMC)\cite{yao1982protocols}, or Homomorphic Encryption (HE) \cite{gentry_fully_2009}. HE allows an untrusted third party to perform certain operations directly on encrypted data, thus,  preserving the data confidentiality. Like many of the works \cite{dowlin2016cryptonets, hesamifard2017cryptodl, cheetah, juvekar2018gazelle, aharoni2023helayers} in PPML literature, we focus on the HE-based inference of ML models trained on plaintext data, which we refer to as Private Inference (PI).
Many PI solutions rely on HE and/or Secure Multi-party Computation (SMC) protocols like Garbled Circuits (GC) \cite{bellare2012foundations}, Oblivious Transfer (OT) \cite{brassard1987all} or Secret Sharing (SS) \cite{shamir1979share}. The approaches that require client interaction are called interactive \cite{cheetah, juvekar2018gazelle} while the others without client's aid are called non-interactive, which typically only use HE \cite{lou2021hemet,aharoni2022he, cai2022hunter, aharoni2023helayers,hesamifard2017cryptodl}. The interactive approaches involve the client in some part of the computation, e.g., non-polynomial functions that HE can not evaluate. Although these approaches could perform a broader range of functions with the client's aid and achieve comparable accuracy to plaintext ML models, they require the client to remain online and involve multiple communication rounds with the cloud throughout the computation, resulting in high communication costs and extended runtime \cite{cheetah}. Additionally, they might be vulnerable to model-extraction attacks \cite{lehmkuhl2021muse}. To this end, here we focus on non-interactive PI, where the entire computation is performed under encryption without client interaction; however, our approach could also be applied to interactive solutions. Since the non-interactive solutions are restricted to the supported operations of HE, typically addition and multiplication, they can only evaluate polynomial functions. So, the layers of the ML models with non-polynomial computations, like activation or pooling layers, should be substituted by approximate polynomial functions to be evaluated using HE operations. After this conversion of the model, we call it HE-friendly. MOFHEI performs this step on the given trained model using our learning-based method automatically and optimizes the model accordingly to keep its performance.\\ 
Despite advances in HE schemes, either in theory or implementation, in the last decade, HE operations are still orders of magnitude slower than their plaintext counterparts and impose substantial memory requirements \cite{aharoni2022complex}. As the model's depth and size grow, this performance gap between plaintext and HE computations increases substantially. To narrow this gap, we need to optimize the trained model. One of the practical solutions in the plaintext domain to minimize the computation overhead is to reduce the model's redundant parameters \cite{han2015deep,blalock2020state}. ML models can be over-parameterized, consisting of parameters that do not necessarily contribute to the model's performance \cite{neyshabur2017exploring,liu2017learning}. Multiple works \cite{han2015deep, frankle2018lottery, joseph2020programmable, zhu2017prune} show that it is possible to ``prune'' these model parameters for more efficient computation without causing a significant loss in accuracy. In the ML community, model pruning refers to selecting a set of parameters based on a specific metric, e.g., parameters with a magnitude below a threshold, setting them to zero, and then fine-tuning the pruned model to recover accuracy loss \cite{han2015learning}. However, recent studies \cite{aharoni2022he, cai2022hunter} demonstrate that applying conventional pruning methods to the encrypted domain offers minor to no benefits in reducing the overhead in HE computations. The underlying reason is that these methods do not consider the way operations are performed in the HE domain. Most HE schemes like Cheon-Kim-Kim-Song (CKKS) \cite{cheon_homomorphic_2017} and Brakerski/Fan-Vercauteren (BFV) \cite{smart2010fully} support single-instruction multiple-data (SIMD) \cite{smart_fully_2014} technique in order to reduce HE computational cost in which a set of values are packed into 'slots' of one ciphertext/plaintext. As an operation is called upon it, this operation is performed simultaneously on all values in all slots at no additional cost. Since existing pruning methods do not account for SIMD packing requirements, they prune model parameters at arbitrary positions within the weight matrix. This random pruning does not substantially reduce the number of expensive HE operations that are responsible for the high latency and memory demands of HE computations. The reason is that as long as non-zero elements reside in packed operands of HE operations, corresponding HE operations can not be eliminated, so there would not be any gains by skipping any HE operation \cite{cai2022hunter,aharoni2022he}. For example, the HE multiplication operation between one ciphertext containing encrypted input data and one plaintext with vectorized weight values can be skipped during PI if and only if all the elements in the plaintext pack are all-zero (pruned). To rectify this issue, we propose an iterative block pruning method that aligns with the data packing method. This approach significantly reduces the number of HE operations, resulting in faster and more efficient PI. It is important to note that while we use batch packing \cite{dowlin2016cryptonets}, our proposed pruning method is versatile and can be applied to other data packing techniques as well.
In this paper, we make the following contributions: 
\begin{enumerate} 
\item We propose a learning-based approach to convert given pre-trained model to its HE-friendly version. It iteratively (1) learns the coefficients of approximate polynomial functions replaced by activation layers, (2) replaces max-pooling layers with HE-compatible average-pooling layers, and fine-tunes the model to maintain its performance.
\item We introduce an iterative block pruning method that considers the data packing method in the encrypted domain and prunes the model accordingly, significantly reducing the number of required HE operations, thus achieving a faster and more cost-effective PI.
\item We evaluate the efficiency of our pruning technique for PI through comprehensive experiments using CKKS \cite{cheon_full_2018} scheme over various model architectures, e.g., LeNet, Autoencoders on MNIST \cite{lecun-mnisthandwrittendigit-2010}, CIFAR-10 \cite{krizhevsky2009learning}, Chest X-Ray \cite{wang2017chestx}, and Electrical Grid Stability Simulated (EGSS) \cite{arzamasov2018towards} datasets. Our results demonstrate that not only does our method reduce PI latency and memory footprint through dropping a considerable number of HE operations, but it also introduces no to minor accuracy loss even at high pruning ratios.
\item We provide MOFHEI, a model optimizing framework that facilitates and enhances PI by making the model HE-friendly automatically and performing an iterative pruning method on it, bringing about more practical applicability in PI. The source code is available at \href{https://github.com/inspire-lab/MOFHEI}{\textit{https://github.com/inspire-lab/MOFHEI}}
\end{enumerate}

The rest of the paper is organized as follows: In section \ref{sec:related}, we discuss the recent state-of-the-art practices contributing to efficient PI. In section \ref{sec:back}, we provide necessary background information about HE, packing methods in HE, and our threat model. Section \ref{section:method} thoroughly explains our proposed framework. We describe our experiments, datasets, hyperparameters, security measures, and resources we utilized in our experiments in section\ref{section:setup}. In section \ref{section:evaluation}, we evaluate the experimental results, compare our framework with the state-of-the-art, and discuss strengths and limitations as well as future directions. Section \ref{section:conclusion} concludes the paper.

\section{Related Works}
\label{sec:related}
Multiple techniques have been employed in the PPML literature to make PI more efficient and practical;
Downlin et al. \cite{dowlin2016cryptonets} present CryptoNets as one of the first works to make non-interactive PI practical, which replaces max pooling with scaled average pooling and activation functions with square functions to make model HE-Friendly and reducing network depth by merging consecutive linear layers. They also proposed the batch packing method we use in our work (More details in \ref{section:packedHE}). 
Hesamifard et al. \cite{hesamifard2017cryptodl} develop CryptoDL, a system akin to CryptoNets, to identify improved low-degree polynomial approximations for activation functions using Chebyshev polynomials \cite{rivlin_chebyshev_2020} and employ the BGV FHE\cite{brakerski2014leveled} scheme with a constant scaling factor to address the issue of large intermediate values\\
Another line of research is Neural Architecture Search (NAS), which strives to find an optimized PPML architecture considering the cryptographic operations and reduce the number of complex functions like ReLU. Although our approach is orthogonal to this line of research, both try to reduce the number of required crypto operations to evaluate the model.
Ghodsi et al. \cite{ghodsi2020cryptonas} propose CryptoNAS, which aims to define a ReLU budget for PI tasks and identify the most efficient networks for a given ReLU budget.

Quantization is another technique used to enhance the computational efficiency of models for PI. While pruning optimizes the model by removing weights and connections, quantization enforces specific values on the weights. Jacob et al. \cite{jacob2018quantization} introduces a quantization approach for an on-device inference that employs integer-only computation, which is more efficient than floating-point inference. They also develop a method to maintain model accuracy after quantization. 
In Faster CryptoNets, Chou et al. \cite{chou_faster_2018} minimize the number of multiplications in a network by pruning individual weights, using the approach from \cite{guo2016dynamic}, and quantizing the remaining weights to increase the sparsity in the weights' polynomial encoding. Using weights that are a power of two, the authors can employ a more efficient multiplication algorithm \cite{akleylek_efficient_2016}.
Podschwadt et al.\cite{podschwadt2024memory} propose a dynamic loading and caching mechanism of recurring values in the repeating structure of convolutional neural network (CNN)s to reduce memory costs.\\
A few recent works propose pruning methods that take HE data packing into account to boost the PI. Hunter by Cai et al. \cite{cai2022hunter} proposes pruning diagonal weight values packed as plaintext vectors to skip the permutation operations required to sum up the intermediate multiplication results in the dot product function. They adopt the packing method of GAZELLE \cite{juvekar2018gazelle}, an SMC-based PI solution, and delegate computation of activation functions to clients. Our work, however, does not rely on the client and approximates activation layers using our proposed learning-based method. So, we can evaluate activation functions under encryption on the server side. Furthermore, Hunter's pruning time for LeNet on the MNIST dataset is over one hour, while ours is in the scale of a few minutes, even for an over 90 percent pruning ratio. Ran et al. \cite{ran2023spencnn} present a CNN inference framework that introduces an HE-group convolution along with group-interleaved encoding, which optimizes channel location inside ciphertexts and iteratively prunes a whole set of weights in 'sub-blocks' corresponding to an inner-rotation operation, altogether reducing the number of costly HE-rotations, cutting down overall latency. However, their grouping approach and its associated encoding method, and the specific sparsity patterns within convolutional groups, are exclusively designed for multi-channel convolutions, and their applicability for other model layers is not investigated. Our pruning method is more versatile, and we show it effectively works for fully-connected and convolutional layers, and it's not dependent on a specific encoding method.

 Aharoni et al. \cite{aharoni2022he} propose HE-PEx that prunes the model according to the tile-packing \cite{aharoni2011helayers} method in a couple of steps. First, they randomly prune the model weights. Then relocate the remaining weights using permutation operations to gather them in fewer packs and make more all-zero packs to prune, reducing the number of required HE operations. Next, there is an expansion step, which expands weight values in the remaining packs to take the most advantage of them for the model's performance. Their method requires two transformation operations, one on input and one on output data, due to the alterations in network structure during pruning. This extra step imposes additional computation. Our method, however, does not require these two extra transformation operations on input and output data, and permutation-expansion steps after pruning.

\section{Background}
\label{sec:back}
This section presents background information about HE, packing methods in HE, especially batch packing, and HE schemes, and concludes with our presumed threat model.
\subsection{Homomorphic Encryption}
HE is a particular type of encryption that allows for certain computations on encrypted data. The data is never decrypted during computation, and the result will also be encrypted. The result of the computation on encrypted data after decryption is the same as if performed on plain data. This makes HE useful for outsourcing computation to untrusted parties. 
HE schemes can be grouped into categories based on their properties, such as message space, supported operations, and functions they can evaluate \cite{armknecht_guide_2015}. The most powerful schemes are FHE schemes. 

FHE schemes support unlimited addition and multiplication operations thanks to bootstrapping \cite{gentry_fully_2009}, but they are limited in the type of functions they can evaluate. Functions that can not be expressed as a polynomial, e.g., comparison, must be approximated and are often computationally expensive. On top of that, encryption adds noise to conceal the data, which is removed during decryption. Performing operations on encrypted data elevates the noise level, and after passing a certain threshold, correct decryption becomes impossible. Since multiplication causes much higher noise growth, the number of multiplications is the limiting factor. The number of consecutive multiplications that can be evaluated on ciphertext before it can no longer be decrypted is called multiplicative depth. Therefore, multiplicative depth also limits the functions that can be evaluated. Although bootstrapping can circumvent this issue by refreshing noise levels, it is computationally expensive. This can be avoided by using leveled FHE schemes. The multiplicative depth can be configured in these schemes using the crypto parameters. Leveled FHE schemes can be helpful when the multiplicative depth of the function is known before the computation.

\subsection{Data Packing in Homomorphic Encryption}
\label{section:packedHE}
Most HE schemes offer Single Instruction Multiple Data (SIMD) \cite{smart_fully_2014} operations to offset the high resource requirements of HE computation. With SIMD, multiple values can be packed into a ciphertext or plaintext. The number of values in each pack depends on the crypto parameters and, in practice, is between $2^{10}$ and $2^{16}$. We can think of SIMD plain-/ciphertexts as vectors of values. All operations between SIMD plain-/ciphertexts are applied slot-wise to the underlying values at no additional cost. This technique can be optimized to perform certain functions. For example, Brutzkus et al. \cite{brutzkus2019low} and Lee et al. \cite{lee2022privacy} propose SIMD packing techniques for efficient convolutions. Aharoni et al. \cite{aharoni2011helayers} introduce a 'tile tensor' data structure for SIMD packing tensors in arbitrary shape in fixed-size chunks called 'tile', encoded HE ciphertext/plaintext, and present their algorithm for 2D convolution based on it. Dowlin et al. \cite{dowlin2016cryptonets} propose another technique, batching or batch packing. In batch packing, the data is grouped into batches of multiple instances, and the same feature of every instance is packed into a ciphertext. This means the number of ciphertexts is equal to the number of features. Assume a batch of data consisting of $n$ instance $X_1, X_2,...,X_n$ and each instance $X_i; \forall i \in [1,n]$ consisting of $m$ features $ X_i = [x_{i,1},x_{i,2},...,x_{i,m} ]$. Then batch packing creates $m$ ciphertext $c_1, c_2, ...,c_m$ where each $C_i = [x_{1,i}, x_{2,i},...,x_{n,i}];\forall i \in [1,m]$. While this technique requires more ciphertexts (therefore more memory), and higher latency, it provides higher throughput. Additionally, it avoids costly rotation operations. It should be noted that we must encode plain values into plaintext. This encoding is necessary since we can only perform operations between ciphertexts and ciphertexts or ciphertexts and plaintexts. We adopt this packing method in our work.
\subsection{HE Schemes}
Most modern HE schemes are based on the hardness of the Ring Learning With Errors problem. While BGV \cite{brakerski_leveled_2014} and BFV \cite{fan_somewhat_2012} only support integers, and TFHE \cite{chillotti2020tfhe} only supports individual bits and binary gates, the CKKS or (HEAAN) \cite{cheon_homomorphic_2017} scheme we use here, supports approximate computation on real numbers. Since the computation is approximate, the plain and encrypted data results will differ. However, the approximation first appears in the least significant bits and can be controlled with crypto parameters. How much approximation error is acceptable depends on the application.

\subsection{Threat Model}

Our threat model consists of three entities: the model owner ($O$), the cloud server ($S$), and the client ($C$). $O$ owns the trained model, i.e., model architecture and all trained parameters. $S$ performs PI on the trained model using encrypted data. $C$ owns private data, encrypts it, and sends it to the cloud server for PI. $O$ can send the model to $S$ in encrypted or plain format; in the former case, $S$ should not learn anything about the model other than its architecture. $C$ should also learn nothing about the model. In our scenario, $S$ knows the model parameters. We assume that both $S$ and $C$ are honest but curious, i.e., $C$ tries to extract model parameters from inference results; $S$ evaluates $C$'s requested functions without deviation but is also interested in revealing $C$'s confidential data. 

\begin{figure*}[ht!]
     \centering
     \begin{subfigure}{0.3\textwidth}
         \centering
         \includegraphics[width=\textwidth]{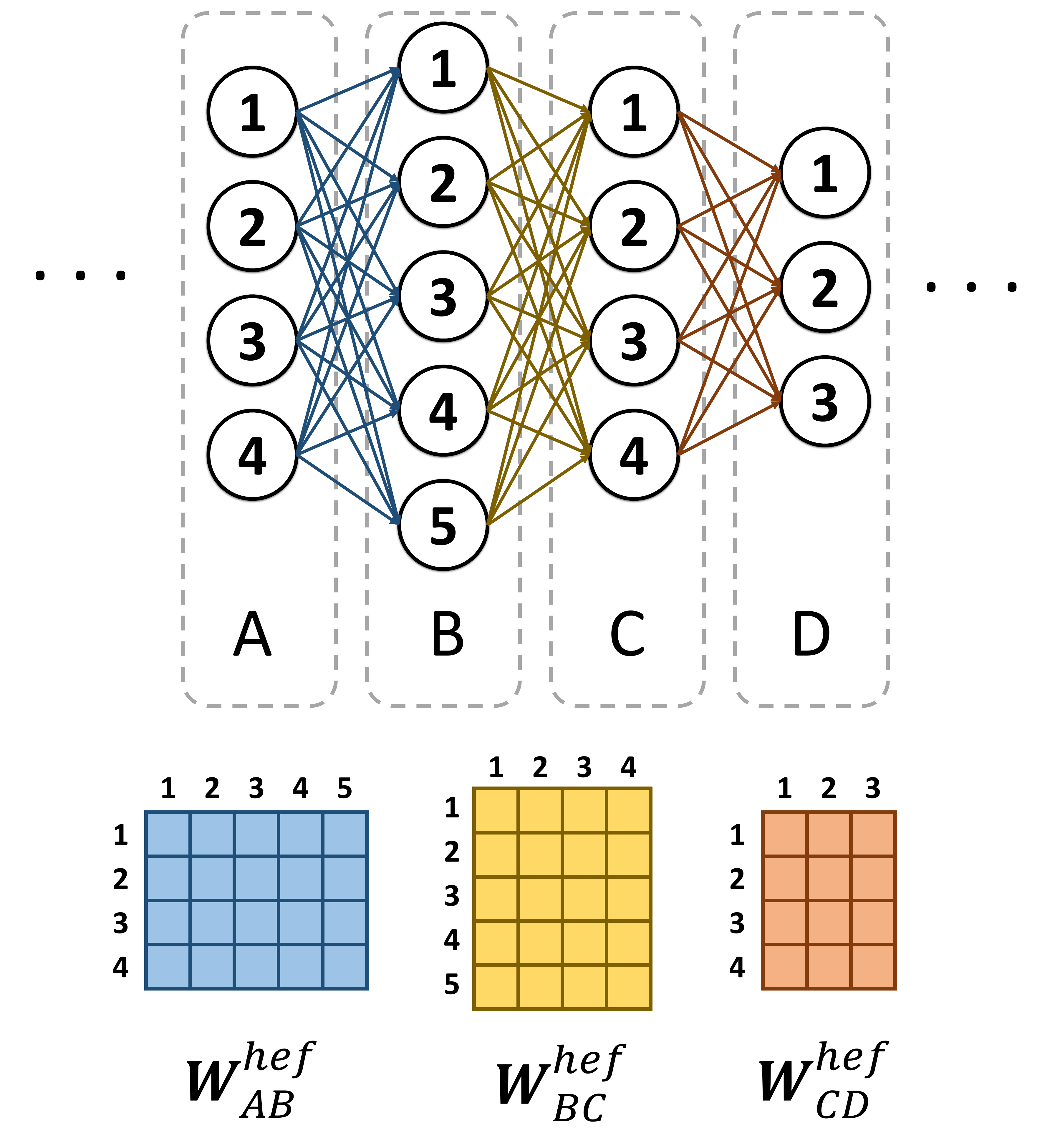}
         \caption{Original/He-Friendly}
         \label{fig:org}
     \end{subfigure}
     \hfill
     \begin{subfigure}{0.6\textwidth}
         \centering
         \includegraphics[width=\textwidth]{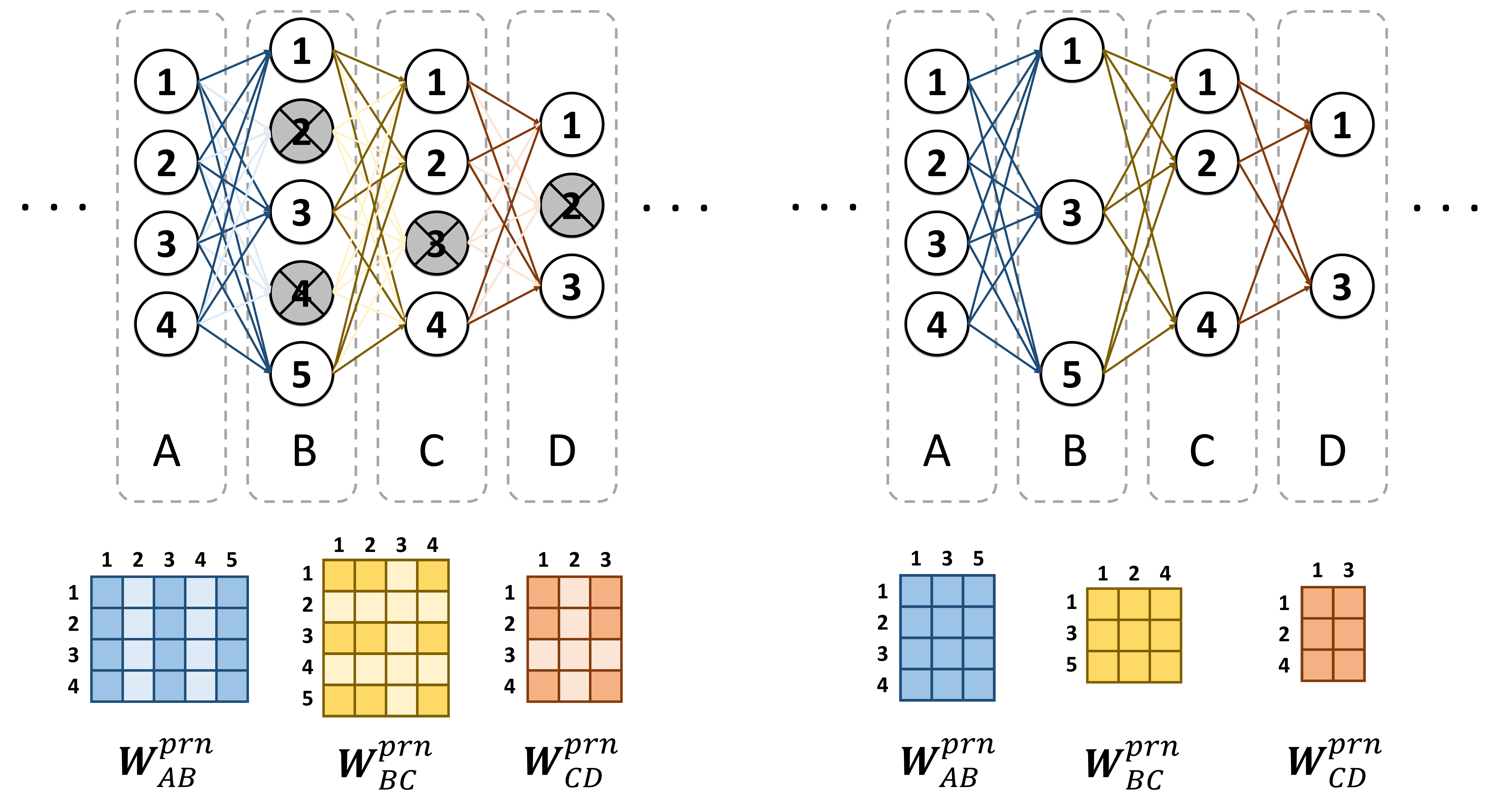}
         \caption{Pruned with all-weights zero columns/rows (left) and with them removed (right)}
         \label{fig:pruned}
     \end{subfigure}
        \caption{An example of HE-friendly fully-connected layer's pruning process with clomun-wise pruning blocks}
        \label{fig: prune dense}
\end{figure*}

\section{The Proposed Framework}
\label{section:method}
We assume the “original model” is a plain-data trained model which can consist of two-dimensional convolution (with or without padding), two-dimensional pooling (max, average, or min – no padding), batch normalization, fully-connected, activation, and/or dropout layers. Our framework has two steps to create the HE-friendly pruned version of the original model with comparable accuracies (steps 1 and 2). Finally, in step 3, we perform PI on this version of the model. We elaborate on these steps in this section:

\begin{figure*}
    \centering
    \includegraphics[width=1\textwidth]{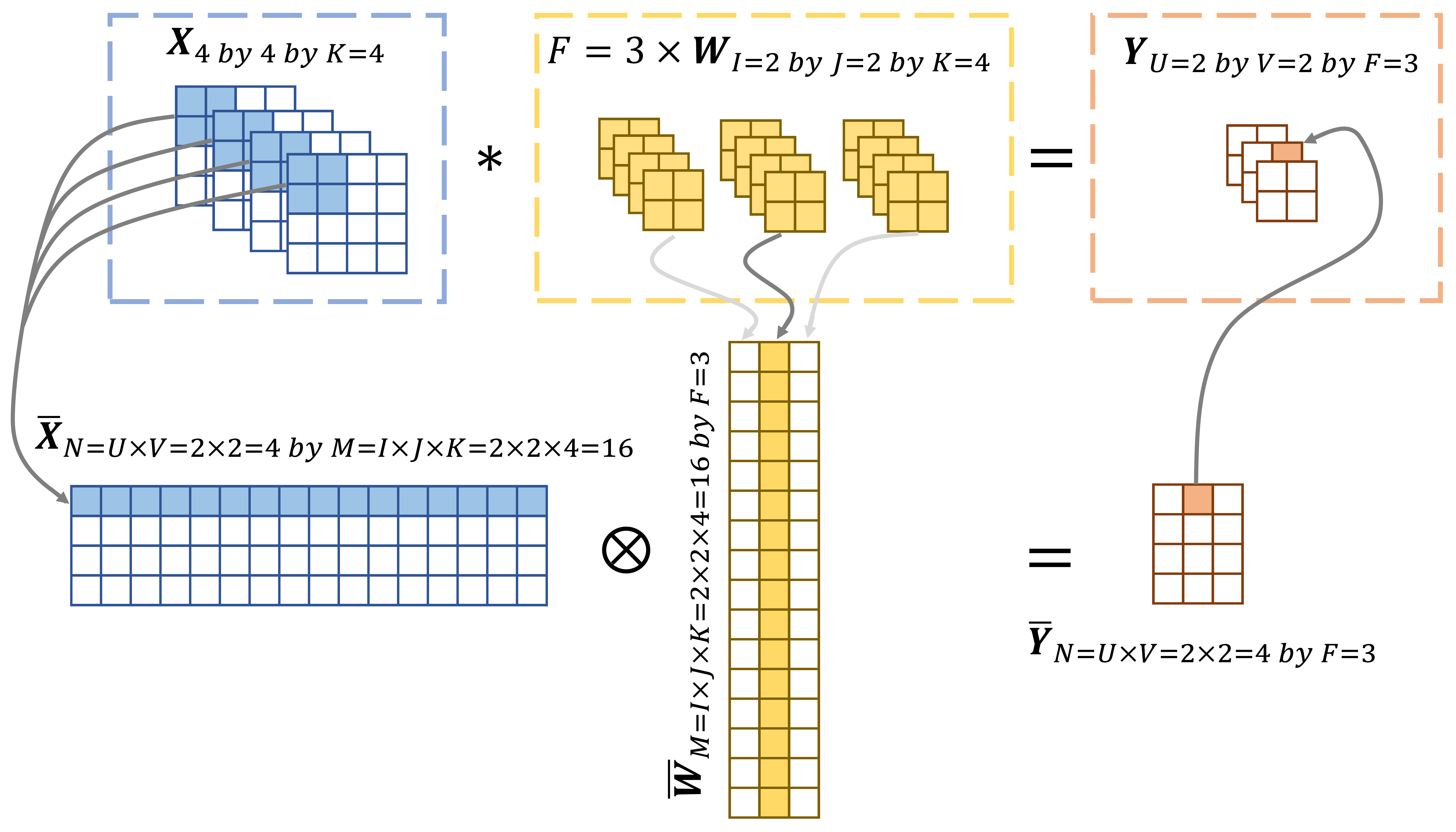}
    \caption{Example of the proposed Conv-Dense operation}
    \label{fig:Conv Transfer}
\end{figure*}

\subsection{Step 1: Learning-based Method to Make an Original Pre-trained Model HE-Friendly}
In this step, we convert all max-pooling layers to average pooling with the same window size and strides as their max-pooling version, one at a time, starting with the latest one (the closest to the output). Accuracy decay of modifying later layers would have a higher recovery chance than earlier ones. Since the incurred error of modifying later layers will be fed forward to fewer subsequent layers, it statistically would produce lower propagated errors in the output layer than the propagated errors incurred by modifying earlier layers. Once we convert a pooling layer, we train only the subsequent layers for a few epochs. Then we fine-tune the whole model with a relatively small learning rate. Next, similar to pooling layers, we convert activation layers. Starting from the latest activation layer, we replace them with a trainable polynomial with a pre-defined degree or a non-trainable square function, one at a time. In the case of polynomials, to obtain the approximate polynomial of the activation function, first, we only train the polynomial coefficients as the trainable parameters of the model for a few epochs, with small magnitude initial weights and a small learning rate. This avoids large loss values. Next, we fine-tune the whole model with a small learning rate. In the square function case, only the fine-tuning step is required. It should be noted that fine-tuning with relatively small learning rates is to avoid overfitting. 

The outcome of step 1 is an HE-friendly model with $A^{\text {hef}} \approx A^{\text {org}}$ where $A^{\text {hef}}$ and $A^{\text {org}}$ are the HE-friendly and original model accuracies, respectively.

\subsection{Step 2. Iterative Block Pruning}
As we discussed in section \ref{sec:Intro}, an effective pruning for HE should consider how data is packed. We need to zero out all values that are going to reside in one pack to skip the computation of that pack. Here, we present a configurable iterative block pruning method that enables the pruning of the model weights in any arbitrary block shape so it matches the way weight values are packed in one plaintext/ciphertext. Zhu et al. \cite{zhu2017prune} propose an over-training pruning method that uses a binary mask to gradually zero out low-magnitude weights until achieving a predefined target sparsity. We extend their method by defining a boundary constraint, which partitions the weight matrix with the predefined shape (height, weight) blocks. We zero-pad the weight matrix if needed. To prune each layer, instead of a binary mask with the size and shape of its weight matrix, we build the binary block mask with the size of blocks placed on the weight matrix. The binary block mask determines which weight blocks participate in the forward pass. We define a pruning schedule, beginning at an initial sparsity value $s_i$ (usually 0) to a final sparsity value $s_f$ throughout $n$ pruning steps, starting at training step $t_0$ and with pruning frequency $\Delta t$. At each pruning step, we sort the blocks based on the average of the absolute value of their weights and mask the smallest blocks to zero up to the desired sparsity level of the pruning step. The block masks are no longer updated once the model achieves the target sparsity $s_f$. In the back-propagation pass, the weights inside the blocks that were masked in the forward pass are excluded from the trainable parameters list and do not get updated. The model recovers from pruning-induced accuracy loss on training steps between each round of block mask update (pruning step). Algorithm \ref{alg:IB} shows the Pseudo code of our proposed Iterative Block Pruning method.
\begin{algorithm}
\caption{Iterative Block Pruning Algorithm}\label{alg:IB}
\textbf{Input:} A Layer, list of Layers, or Model, Block shape, Pruning Schedule \\
\textbf{Output:} Block Pruned Layer, list of Layers, or Model
\begin{algorithmic}
\State \text{Blk\_Masks = \textit{Generate\_block\_masks}(Block Shape, $w_{\text{Layer(s)}}$)}
\State\textbf{For} step in Pruning steps:
\State \hspace*{\algorithmicindent}Blk\_Masks = \textit{Sort\_by\_avg\_block\_weights}(Blk\_Masks)
\State \hspace*{\algorithmicindent}Blk\_Masks = \textit{Zero\_mask\_smallest\_blocks}(Blk\_Masks,
\State \hspace*{\algorithmicindent}Pruning Schedule)
\State \hspace*{\algorithmicindent}\textit{Back\_Propagation}(Blk\_Masks, $w_{\text{Layer(s)}}$)
\end{algorithmic}
\end{algorithm}

We first demonstrate our approach for pruning fully-connected layers and then explain how we apply it to convolution layers by running a transformation procedure on them. Since in this work, we use batch packing (see \ref{section:packedHE}), for each fully-connected layer except the output layer, we select the block shape to be $M$ by $1$ where $M$ is the number of rows in the corresponding weight matrix; this results in pruning columns of the weight matrix. Pruning one entire column of the weight matrix allows skipping $M$ HE multiplications and $M-1$ HE additions.  We highlight that our pruning method is not restricted to column-wise pruning and is capable of pruning weight matrices using any arbitrary block shape. One can match the pruning block shape of our method with any data encoding format used by other packing methods to prune the model efficiently. Once the model is pruned, every prunable layer consists of a set of all-zero weight columns depending on the predefined sparsity. To facilitate the computations on the HE side, we remove the all-zero columns from the weight matrix, which, in fully-connected layers, is equivalent to removing neurons from the layer. Similarly, any accuracy decay induced by dropping pruned columns, e.g., removing biases, is recovered by fine-tuning the remaining parameters. It should be noted that removing neurons from a fully-connected layer affects the shape of its output feature map. All rows of the following layer's weight matrix corresponding to the all-zero columns of the pruned layer's weights need to be removed (see Figure \ref{fig:pruned}) to keep dimensions consistent. This explains why the sparsity of the pruned model is noticeably higher than the predefined sparsity of each prunable layer. Figure \ref{fig: prune dense} illustrates the steps of our pruning method (column-wise) for fully-connected layers schematically.


\begin{figure}[ht!]
    \centering
    \includegraphics[width=0.45\textwidth]{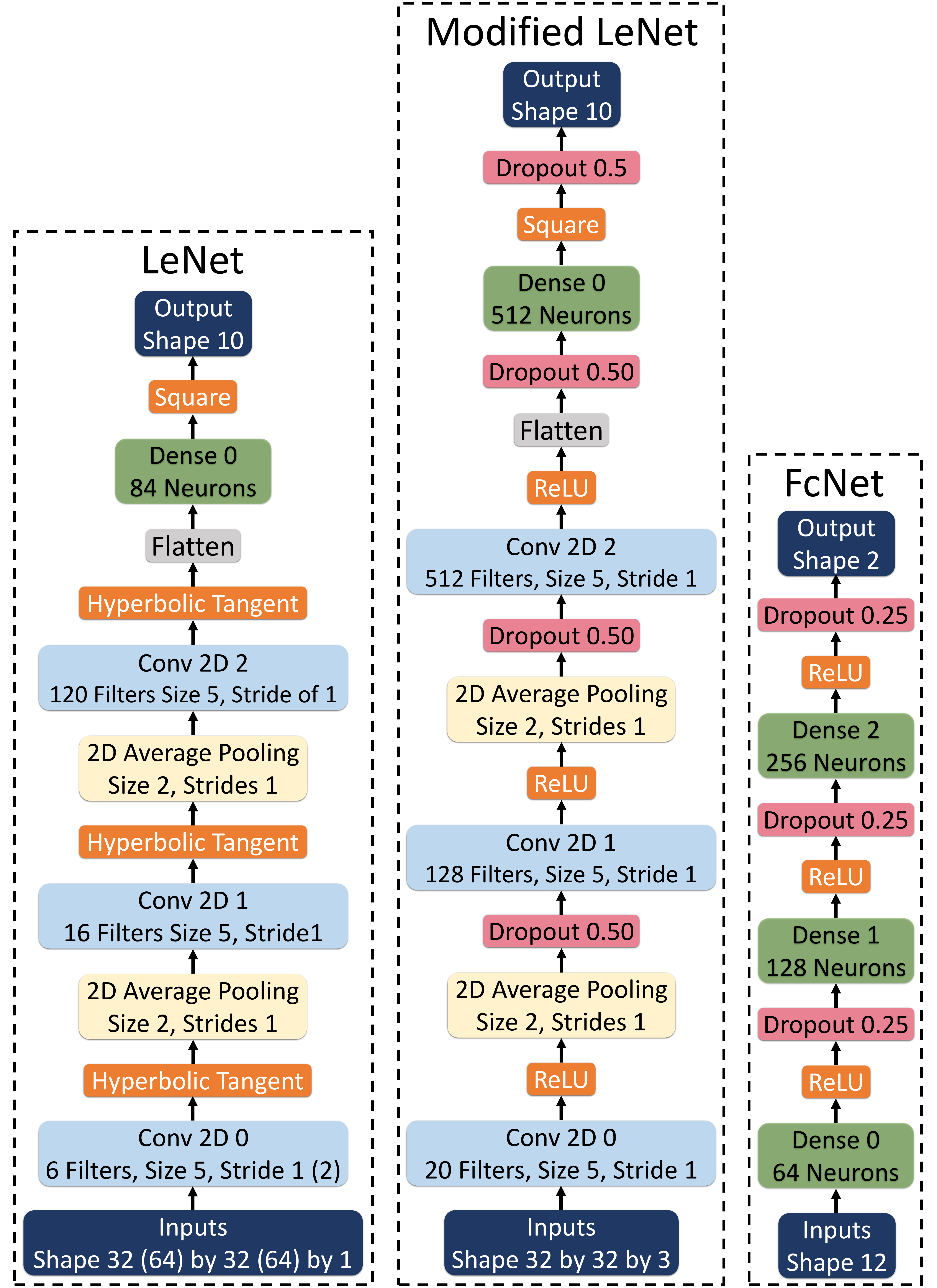}
    \caption{Classification model architectures; the values inside the parenthesis represent slight modifications for X-Ray dataset}
    \label{fig: model classification}
\end{figure}

\begin{figure}[ht!]
    \centering
    \includegraphics[width=0.3\textwidth]{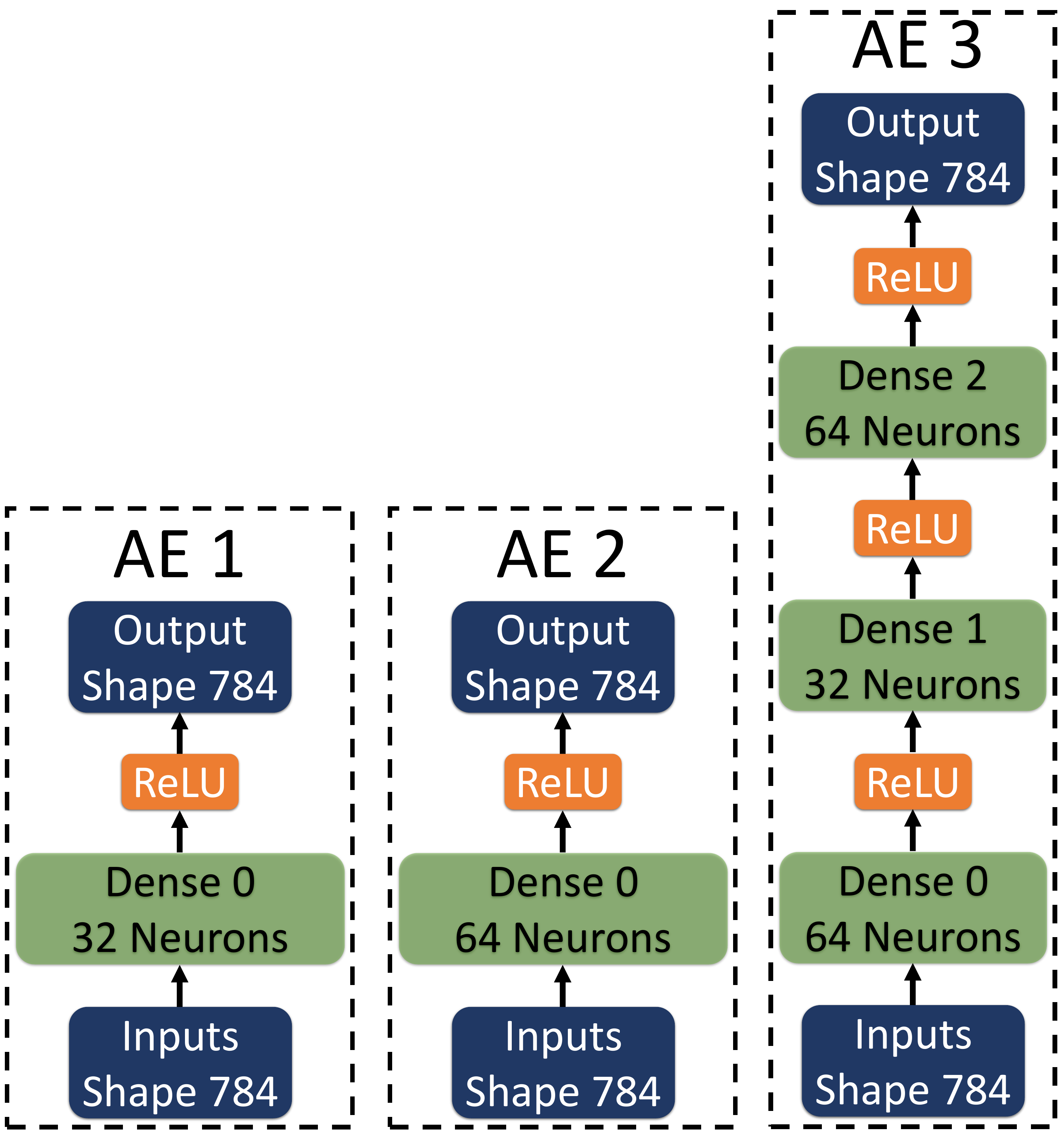}
    \caption{Regression/Autoencoder (AE) model architectures of Aharoni et al. \cite{aharoni2022he} (HE-PEx)}
    \label{fig: model regression}
\end{figure}

 In order to benefit from our iterative block pruning method for convolution layers, we transform a two-dimensional convolution layer into an equivalent fully-connected layer using a transformation technique explained below, referred to as the Conv-Dense operation in this paper. Suppose a two-dimensional convolution layer with $F$ filters of the size $I$ by $J$ by $K$ in weight matrix $\mathbf{W}$ and $F$-dimensional bias vector $\mathbf{B}$ is to map a K-channel feature map, $\mathbf{X}$, into a $U$ by $V$ by $F$ feature map, $\mathbf{Y}$. For every stride, every $I$ by $J$ by $K$ filter is multiplied by a same size volume chunk in  $\mathbf{X}$ elementwise to produce a new volume; all values of the new volume are summed with each other along with the filter bias to produce the corresponding value in $\mathbf{Y}$. Since the multiplication is elementwise, the volume chunk and the filter can be reshaped (i.e., flattened) into a row and a column vector, respectively, with the shapes 1 by $M = I \times J \times K$ and $M$ by 1. Next, they can be multiplied by each other using a dot product and added to the bias. This is the primary idea for converting two-dimensional convolutions into their equivalent fully-connected representation. Given the strides in every direction, one can extract all volume chunks from $\mathbf{X}$ and broadcast them into $\mathbf{\bar{X}}$ with the size  $N = U \times V $ by $M$ where each row represents a volume chunk in $\mathbf{X}$ corresponding to a different stride. Similarly, all filters in matrix $\mathbf{W}$ can be reshaped into $\mathbf{\bar{W}}$ with the size $M$ by $F$ where each column corresponds to a different filter in $\mathbf{W}$. The two-dimensional representation of the $\mathbf{Y}$ is the $N$ by $F$ matrix $\mathbf{\bar{Y}} = \mathbf{X} \otimes \mathbf{W} + \mathbf{B}$ ($\otimes $ represents dot product) which is the equivalent computation in a fully-connected layer; $\mathbf{\bar{Y}}$ then can be easily reshaped into $\mathbf{Y}$. With this transformation technique, we can apply our iterative block pruning method to a convolution layer. Pruning columns of $\mathbf{\bar{W}}$ in the equivalent fully-connected representation correspond to pruning filters of the transformed convolution layer. Therefore, removing pruned columns of $\mathbf{\bar{W}}$ actually drops corresponding filters from the convolution layer. Finally, we perform another round of fine-tuning to compensate for the accuracy loss due to dropping filters. Figure \ref{fig:Conv Transfer} illustrates an example of Conv-Dense operation when $F=3$, $I=2$, $J=2$, $K=4$, $U=2$, $V=2$, $M=2\times 2 \times 4 = 16$, and $N=2 \times 2 = 4$. 

The outcome of step 2 is a HE-friendly pruned model with accuracy $A^{\text {prn}} \approx A^{\text {org}}$.

\begin{table*}[]
\centering
\caption{Experiment settings and configurations}
\label{tab:Experiment settings and configurations}
\small
\begin{tabular}{lllllllllllllll}
\toprule
Experiment       & PM  &  TRS    & VLS   & TES    & OE & TE  & FE  & PE & RE    & LR   & TL   & FL   & PMD   & CM   \\
\midrule
MNIST-LeNet      & CLS &  57.0K & 3.0K & 10.0K  & 100 & 100 & 100 & 10 & 100   & 1e-3 & 1e-3 & 1e-4 & 32768 & 860\\
X-Ray-LeNet      & CLS &  5216  & 16   & 624    & 100 & 100 & 100 & 99 & 100   & 1e-3 & 1e-3 & 1e-4 & 32768 & 860\\
CIFAR-10-MLeNet   & CLS &  47.5K & 2.5K & 10.0K & 100 & 100 & 100 & 50 & 100   & 1e-3 & 1e-3 & 1e-4 & 32768 & 860 \\
EGSS-FcNet       & CLS &  8550  & 450  & 1.0K   & 100 & 100 & 100 & 10 & 100   & 1e-3 & 1e-3 & 1e-4 & 16384 & 290\\
MNIST-AE1        & REG &  57.0K & 3.0K & 10.0K  & 20  & 50  & 50  & 10 & 5     & 1e-3 & 1e-3 & 1e-4 & 8192  & 200 \\
MNIST-AE2        & REG &  57.0K & 3.0K & 10.0K  & 30  & 50  & 50  & 10 & 10    & 1e-3 & 1e-3 & 1e-4 & 8192  & 200\\
MNIST-AE3        & REG &  57.0K & 3.0K & 10.0K  & 30  & 50  & 50  & 10 & 10    & 1e-3 & 1e-3 & 1e-4 & 16384 & 200\\

\bottomrule
\vspace{-0.5em}
\end{tabular} 

    \begin{tablenotes}{}
    \item 
MLeNet: Modified LeNet;
EGSS: Electrical Grid Stability Simulated dataset;
AE: Autoencoder;
PM: Problem;
CLS: Classification;
REG: Regression;
TRS: Number of Training Samples;
VLS: Number of Validation Samples;
TES: Number of Testing Samples;
OE: Original Model Training Epochs;
TE: Transfer Learning Epochs;
FE: Fine-Tuning Epochs;
PE: Patience Epochs;
RE: Pruning Epochs;
LR: Original Model \& Pruning Learning Rate;
TL: Transfer-Learning Learning Rate;
FL: Fine-Tuning Learning Rate;
PMD: Polynomial Modulus Degree;
CM: Coefficient Modulus (Bits);

    \end{tablenotes}

\end{table*}

\subsection{Step 3. Private Inference}

We take the HE-friendly pruned model of step 2 and extract the weights and layer configurations, to perform PI. We implement the algorithms for PI using the CKKS primitives provided by SEAL \cite{sealcrypto}. Since we use batch packing, we can treat all operations as performed on a single instance (See \ref{section:packedHE}). For example, consider a fully-connected layer with 128 units and 256 inputs. This layer's weight matrix $\mathbf{W}$ has dimensions $128 \times 256$. A batch of inputs $\mathbf{X}$ consists of $b$ instances with $256$ features each, making it a $b \times 256$ matrix. The layer output is computed as $\mathbf{Y\mathbf} = \mathbf{X} \otimes\mathbf{W}^T$ (in this example, we ignore the bias for simplicity), with $\mathbf{Y}$ being $b \times 128$. With batch packing, the input batch, $\mathbf{X}$ is packed into 256 ciphertexts. We can think of these 256 ciphertexts as a vector. The computation of the layer stays the same. However, the input and output are now vectors instead of matrices. This works analogously for all neural network layers that do perform computation along the batch axis. However, for this computation to work, we need to encode each value in the weight matrix into a single plaintext. This encoding takes a value from the weight matrix and repeats it $b$ times before turning it into a plaintext. This leads to a large number of plaintext objects. We can parallelize the computation of the dot product by having multiple jobs running simultaneously. If each job computes one $y_k \in \mathbf{Y}$ we can have 128 parallel jobs. The benefit of this parallelization is that it is lock-free since no jobs concurrently alter the same resource or read from a resource another job alters.

\begin{table}
\centering
\caption{
      Comparison of the original model (ORG) and the HE-friendly model (HEF) in
      terms of: performance metric of either testing accuracy (ACC) or mean squared error
      (MSE), and the time (Time) to transform the 
      original model into an HE-friendly one in seconds. 
      }
\label{tab:org_vs_hef}
\small
\begin{tabular}{llrrr}
\toprule
                     Model & Metric &  ORG &  HEF &  Time \\
\midrule
    MNIST-LeNet     &    ACC &  0.99 & 0.99 &    113 \\
    X-Ray-LeNet     &    ACC &  0.84 & 0.75 &    198 \\
    CIFAR-10-MLeNet &    ACC &  0.76 & 0.75 &    NA \\
    EGSS-FcNet      &    ACC &  0.94 & 0.93 &    33 \\
    MNIST-AE1       &    MSE &  0.02 & 0.02 &    92 \\
    MNIST-AE2       &    MSE &  0.01 & 0.01 &    103 \\
    MNIST-AE3       &    MSE &  0.02 & 0.02 &    212 \\

\bottomrule
\end{tabular}

    \begin{tablenotes}{}
    \item AE: Autoencoder; EGSS: Electrical Grid Stability Simulated Dataset; MLeNet: Modified LeNet; NA: Not Available.
    \end{tablenotes}

\end{table}

\section{Experimental Setup}
\label{section:setup}

To illustrate the efficiency of our proposed method, we have designed seven experiments consisting of four classifications and three autoencoder experiments. Three out of four classification experiments are image classifications and employ LeNets \cite{lecun1998gradient}, and a modified version of it (see Figure \ref{fig: model classification} for more details) for the MNIST \cite{lecun-mnisthandwrittendigit-2010} CIFAR-10 \cite{krizhevsky2009learning}, and Chest X-Ray \cite{wang2017chestx} (resized into 64 by 64 gray-scale) datasets. We chose LeNet as a standard benchmark to assess our proposed method, which is used in other works as well\cite{cai2022hunter}; We made some modifications to LeNet to (1) have a model with more diverse and complex layers, (2) increase the capacity of the model for more complicated CIFAR-10 and X-Ray classification. The last classification experiment is a binary classification that employs a custom-made fully-connected model (FcNet) (see Figure \ref{fig: model classification} for more details) for Electrical Grid Stability Simulated (EGSS) Data \cite{arzamasov2018towards}, a non-image dataset with 10K 14-variable samples. The task is to estimate grid stability (i.e., stable or not). More information about the dataset is available at Arzamasov et al. \cite{arzamasov2018towards}. The choice of this experiment is to (1) show our model benchmarks on a non-image real-life infrastructure management task and (2) the ability of our method to work with custom models. Our last experiments are simple autoencoder (AE) network architectures used in the experiments of recent work by Aharouni et al., \cite{aharoni2022he} (see Figure \ref{fig: model regression} for more details). These autoencoders are used for reconstruction tasks on MNIST. We chose these experiments to (1) compare our proposed method with one of the recent state-of-the-art works (i.e., HE-PEx\cite{aharoni2022he}) and (2) illustrate the effectiveness of our method on a different task, i.e., reconstruction task. In our experiments, we developed our trained original models using the hyperparameters summarized in Table \ref{tab:Experiment settings and configurations}. For each experiment, we prune the HE-friendly model 10 times, each time with a different layer sparsity, from 50\%, 55\%, 60\%, ..., to 95\%. However, due to space limitation, we just report the results of pruning and PI for four sparsity levels in Table \ref{tab:giant}.
We normalize 8-bit integer images of MNIST and CIFAR-10 datasets between 0 and 1 by dividing the pixel magnitudes by 255. For the EGSS dataset, we linearly scale the variables between 0 and 1. For all experiments, the model training's (or fine-tuning) stopping criteria is either (1) reaching the maximum corresponding epochs or (2) no improvement in validation metrics (i.e., loss or accuracy) after a predefined number of epochs (i.e., patience epochs), whichever comes first. Furthermore, for a smoother convergence, a 50\% reduction in learning rate is scheduled every five epochs if the validation metrics do not improve with a minimum learning rate of 1e-7. For PI, we adopt batch packing \cite{dowlin2016cryptonets} to encode and encrypt data and use the CKKS scheme of SEAL \cite{sealcrypto} with 128-bit security in all experiments. We utilize the TensorFlow Model Optimization toolkit \cite{tfmot-sparsity-prune-low-magnitude} to implement our iterative block pruning method. We use one Tesla V100-SXM2 with 32 GB memory to train and fine-tune models during conversion to the HE-friendly version, and pruning steps. For PI, we use two 52-core Intel(R) Xeon(R) Gold 6230R CPUs @ 2.10GHz with a total of 755 GB RAM.

\input{giant_table}

\begin{figure*}
    \centering
    \includegraphics[width=\textwidth]{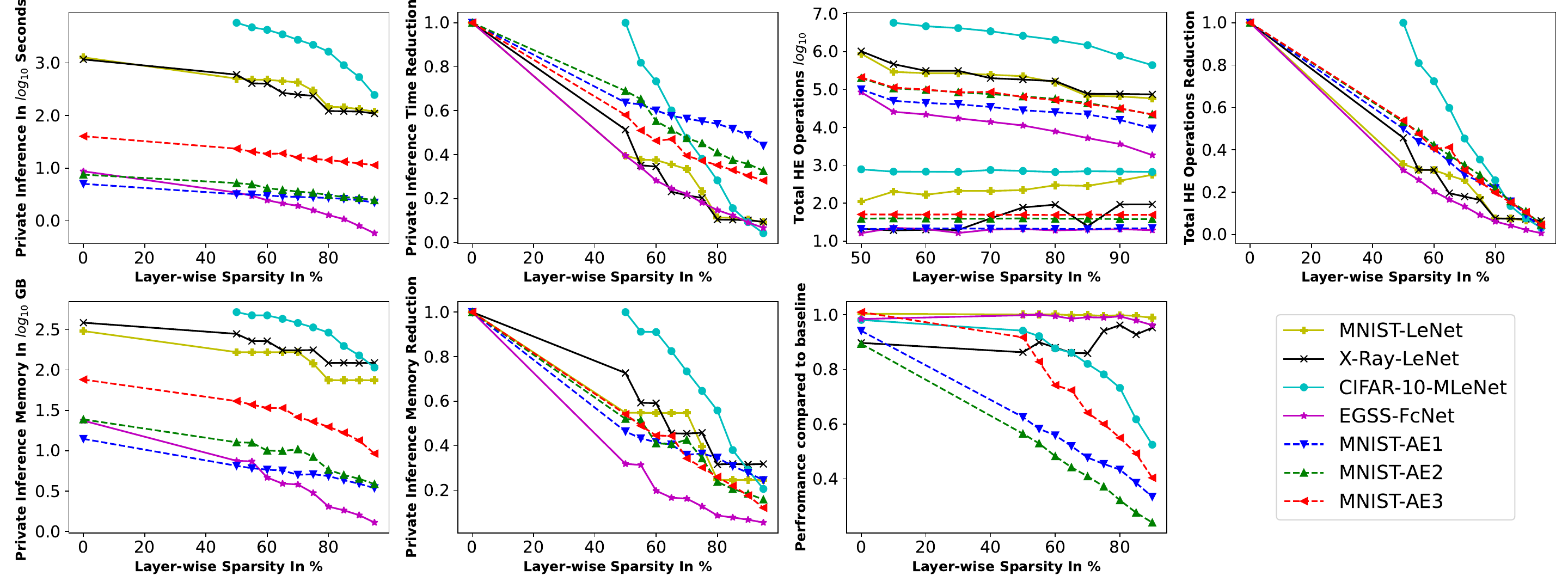}
    \caption{Various performance over layer-wise sparsity. For performance, we use the original model as the baseline.}
    \label{fig:all_plots}
\end{figure*}

\section{Evaluation and Discussion}
\label{section:evaluation}
We first examine the performance of our learning-based method and then discuss the results of each experiment to evaluate our iterative block pruning method. To provide a point of comparison with state-of-the-art works, we compare the results of the MNIST-LeNet experiment with the Hunter paper \cite{cai2022hunter} and the results of the autoencoder experiments with the HE-PEx paper \cite{aharoni2022he}.\\
Table \ref{tab:org_vs_hef} compares the original and HE-friendly versions of our experiments' models trained on various datasets to illustrate the performance of our proposed learning-based method in converting a trained model to its HE-friendly version. As demonstrated, in most cases, our proposed method does not cause a considerable reduction in the models' accuracies, showing the effectiveness of our method. On top of that, as the results show, our method performs the conversion of the model in a reasonable time.\\
Table \ref{tab:giant} presents summaries of our seven experiments results for the HE-friendly model and four out of ten pruned models with layer-wise sparsities of 50\%, 60\%, 80\%, and 90\%. As we stated in section \ref{sec:Intro}, our method's primary goal is to accelerate PI by reducing the number of HE operations. Our experiments' results illustrate the effectiveness of our method in achieving this goal by taking the HE-friendly model as the baseline and comparing it with pruned models at different sparsity levels in terms of the number of HE operations and other factors, including pruning time, maximum required memory, PI latency, accuracy or mean-squared error, and the final sparsity of the model after the pruning. Furthermore, for each experiment model's prunable layer (with regards to notations of Figures \ref{fig: model classification} and \ref{fig: model regression}), we report the number of neurons or filters, the reduction rate of neurons or filters, and the number of HE operations. It should be noted that the total number of HE operations is not equal to the sum of the prunable layer's HE operations; there are non-prunable layers (e.g., poolings, activations, outputs, etc.) in each model with HE operations. Furthermore, for each of our seven experiments and all ten layer-wise sparsities of 50\%, 55\%, ..., and 95\%, Figure \ref{fig:all_plots} schematically presents layer-wise sparsity percentage versus (1) PI latency (in $log_{10}$), (2) PI time reduction rate as compared with HE-friendly model, (3) total HE operations (in $log_{10}$), (4) total HE operations reduction rate, (5) PI memory usage (in $log_{10}$), (6) PI memory reduction rate, and (7) performance (i.e., accuracy or mean-squared error) rate as compared to the baseline (i.e., original model).\\
Comparing the results of our MNIST-LeNet experiment with the results of Hunter \cite{cai2022hunter}, with almost no accuracy drop, shows our method reduces the total number of HE operations by 69.32\% while Hunter \cite{cai2022hunter} achieves only a reduction of 47.35\%. At the highest sparsity level and only a 2\% accuracy drop, our method lowers the number of HE operations by 93.41\%. It brings about 9.63 and 4.04 times reduction in PI runtime and required memory. We should also highlight that it takes 1.2 hours for Hunter's method \cite{cai2022hunter} to prune this model, while our pruning method is performed over-training and is in the scale of a few minutes. Moreover, they perform PI interactively, enabling them to use smaller crypto parameters with lower memory requirements and evaluate deeper and more complex networks in their experiments at the expense of communications costs. However, our model-optimizing framework yields an optimized version of the model, enabling us to perform PI efficiently under encryption without the need for client interaction.

In X-Ray-LeNet experiments, the pruned model, with almost 98\% overall sparsity, not only maintains its performance, but surprisingly it even outperforms the baseline model, indicating that the LeNet could be significantly over-parametrized for the X-Ray dataset. At the highest sparsity, 93.42\% of HE operations are removed, which reduces the runtime and the required memory of PI by 89.62\% and 74.25\%, respectively, in comparison with the baseline model.

Given our resources and the selected crypto parameters, the CIFAR-10-MLeNet (i.e., modified LeNet) HE-friendly (i.e., unpruned) model is too large for us to compute. The first convolutional layer alone requires over 600 GB of memory. Without pruning, we are not able to run the model at all. However, as illustrated in Table \ref{tab:giant}, when this model is pruned up to 73\%, its accuracy loss is still negligible. This enables us to perform PI on the pruned version of the CIFAR-10-MLeNet without considerable loss of performance. We designed this experiment to illustrate that our pruning method enables us to perform PI on more complex models with comparable performance that we wouldn't have been able to compute given resource limitations, making PI more practical.

In the EGSS-FcNet experiment with a final sparsity of 94\%, the pruned model maintained the same performance while 93.85\% of HE operations were dropped. Accordingly, the required memory and PI latency decreased 12 and 8 times, respectively. At the highest level of sparsity, with just a 2\% drop in accuracy, only 2.88\% of the HE operations from the baseline model were retained, demonstrating the pruning method's remarkable effectiveness in significantly reducing HE operations while preserving the model's performance.

In all autoencoders, increasing sparsity leads to performance degradation due to their simple architectures. However, according to Table \ref{tab:giant} and Figure \ref{fig:all_plots}, AE3 performs better than the other two; it keeps its performance up to 50\% layer-wise sparsity. In comparing the results of autoencoder experiments with HE-PEx\cite{aharoni2022he}, we observe that at the almost same reconstruction mean-square error ($\approx$0.03), our method reduces PI latency of AE3 by 65\%, which is the same as their best result on AE3 model. However, our method outperforms in reducing memory consumption and decreases it by 73.68\%, while HE-PEx\cite{aharoni2022he} reduction rate was 65\%. Moreover, \cite{aharoni2022he}'s method needs to perform a set of permutation and expansion operations on the pruned model to align it with packing requirements. To match these changes in the model structure, they have to perform two transformation operations on the input and output of the model, which brings extra complexity. However, our approach prunes the model in the desired block shape during training in alignment with the packing requirements, so there is no need to perform additional post-pruning procedures on the model or any transformations on input and output data. Although they claim their method can be extended to convolutional layers, they just evaluate their method on simple, fully-connected networks (autoencoders). However, we show that our method works on the convolutional layers as well effectively through an extensive set of experiments.

\section{Conclusion}
\label{section:conclusion}
This study presents a model optimization framework, MOFHEI, that optimizes pre-trained ML models for faster and more efficient non-interactive PI under HE. Our approach effectively transforms an ML model into an HE-friendly version and prunes it with respect to the HE packing method. Therefore, while maintaining accuracy, it reduces the number of HE operations, thereby PI latency and memory usage. Our iterative block pruning method works for convolution and fully connected layers, along with batch packing. In our experiments, we demonstrate the effectiveness of our pruning technique by showing considerable differences in the number of HE operations, latency, and memory consumption before and after applying our method in different sparsities. In comparison with state-of-the-art works, MOFHEI has the following advantages: has lower pruning time due to an over-training approach, can optimize complex networks and enable performing PI, offers a learning-based method to convert the model to an HE-friendly version automatically, can be integrated with other packing methods, does not need client interaction and its communication costs, does not need to perform any transformation on input and output data and post-pruning operations,e.g., permutation to match with data packing requirements. The proposed framework contributes to PPML by offering a solution that balances model performance, privacy, and computational efficiency. In future work, we plan to extend the supported layers, such as recurrent neural networks, and integrate our pruning method with different packing methods in addition to batch packing.

\bibliographystyle{IEEEtranS}
\bibliography{IEEEabrv,references}

\end{document}

%% file: giant_table.tex
\begin{table*}
\centering
\caption{Experiments (E) using the HE-friendly (HEF) models
and models with different layer-wise sparsities (columns). We show the final sparsity (Sparsity) of the model, the prunable layers and their 
units (filters) / reduction factor / number HE operations, test accuracy 
(ACC) or mean squared error (MSE), the time need to prune (TP) in seconds,
private inference (PI) latency (TPI) in seconds, and total number of HE 
operations (HEO) / maximum memory required to perform PI (MPI) in GB.
}
\label{tab:giant}
\normalsize
\begin{tabular}{c|lrrrrrr}
\toprule
E & Information &                                                       HEF &              50\% &              60\% &              80\% &              90\% \\
\midrule
\multirow{9}{*}{\rotatebox{90}{MNIST-LeNet}} & Sparsity  &                - &              0.73 &              0.81 &              0.96 &             0.98 \\
 & Conv 0    &    6 / - / 2.4e5 &   3 / 2.0 / 1.2e5 &   3 / 2.0 / 1.2e5 &    1 / 6.0 / 3.9e4 &  1 / 6.0 / 3.9e4 \\
 & Conv 1    &   16 / - / 4.8e5 &   8 / 2.0 / 1.2e5 &   7 / 2.3 / 1.1e5 &    3 / 5.3 / 1.5e4 &  2 / 8.0 / 1.0e4 \\
 & Conv 2    &  120 / - / 9.6e4 &  61 / 2.0 / 2.4e4 &  51 / 2.4 / 1.8e4 &    25 / 4.8 / 3750 &  16 / 7.5 / 1600 \\
 & Dense 0   &   84 / - / 2.0e4 &   48 / 1.8 / 5856 &   39 / 2.2 / 3978 &     18 / 4.7 / 900 &    9 / 9.3 / 288 \\
 & ACC       &             0.99 &              0.99 &              0.99 &               0.98 &             0.97 \\
 & TP /TPI   &         - / 1272 &         111 / 502 &         167 / 477 &          298 / 145 &        393 / 132 \\
 & HEO / MPI &      8.8e5 / 303 &       2.9e5 / 166 &       2.7e5 / 166 &         6.7e4 / 75 &       5.8e4 / 75 \\
\hline
\multirow{9}{*}{\rotatebox{90}{X-Ray-LeNet}} & Sparsity  &                - &              0.70 &              0.79 &          0.96 &              0.98 \\
 & Conv 0    &    6 / - / 2.7e5 &   4 / 1.5 / 1.8e5 &   3 / 2.0 / 1.4e5 &   1 / 6.0 / 4.5e4 &   1 / 6.0 / 4.5e4 \\
 & Conv 1    &   16 / - / 5.8e5 &   9 / 1.8 / 2.2e5 &   7 / 2.3 / 1.3e5 &   3 / 5.3 / 1.8e4 &   3 / 5.3 / 1.8e4 \\
 & Conv 2    &  120 / - / 9.6e4 &  62 / 1.9 / 2.8e4 &  55 / 2.2 / 1.9e4 &  24 / 5.0 / 3600 &  12 / 10.0 / 1800 \\
 & Dense 0   &   84 / - / 2.0e4 &   50 / 1.7 / 6200 &   40 / 2.1 / 4400 &   26 / 3.2 / 1248 &    13 / 6.5 / 312 \\
 & ACC       &             0.75 &              0.72 &              0.74 &              0.81 &              0.80 \\
 & TP / TPI  &         - / 1163 &          21 / 597 &          20 / 401 &          91 / 121 &           93 / 118 \\
 & HEO / MPI &      1.0e6 / 386 &       4.7e5 / 280 &       3.1e5 / 228 &       7.7e4 / 122 &       7.4e4 / 122 \\
\hline
\multirow{9}{*}{\rotatebox{90}{CIFAR-10-MLeNet}} & Sparsity  &              - &               0.73 &               0.83 &                             0.96 &               0.99 \\
 & Conv 0    &   20 / - / N/A &   10 / 2.0 / 1.2e6 &    9 / 2.2 / 1.1e6 &    5 / 4.0 / 5.9e5 &   2 / 10.0 / 2.4e5 \\
 & Conv 1    &  128 / - / N/A &   69 / 1.9 / 3.5e6 &   53 / 2.4 / 2.4e6 &   27 / 4.7 / 6.8e5 &   14 / 9.1 / 1.4e5 \\
 & Conv 2    &  512 / - / N/A &  256 / 2.0 / 8.8e5 &  205 / 2.5 / 5.4e5 &  103 / 5.0 / 1.4e5 &  51 / 10.0 / 3.6e4 \\
 & Dense 0   &  512 / - / N/A &  262 / 2.0 / 1.3e5 &  211 / 2.4 / 8.7e4 &  103 / 5.0 / 2.1e4 &    54 / 9.5 / 5508 \\
 & ACC       &           0.75 &               0.72 &               0.67 &               0.56 &               0.40 \\
 & TP / TPI  &        - / N/A &         781 / 5781 &          678 /4241 &         663 / 1635 &          685 / 537 \\
 & HEO / MPI &      N/A / N/A &        5.8e6 / 521 &        4.2e6 / 474 &        1.5e6 / 291 &        4.4e5 / 152 \\
\hline
\multirow{8}{*}{\rotatebox{90}{EGSS-FcNet}} & Sparsity  &                - &               0.70 &               0.80 &             0.94 &            0.98 \\
 & Dense 0   &    64 / - / 1536 &    45 / 1.4 / 1080 &     37 / 1.7 / 888 &   22 / 2.9 / 528 &  12 / 5.3 / 288 \\
 & Dense 1   &  128 / - / 1.6e4 &    68 / 1.9 / 6120 &    56 / 2.3 / 4144 &  28 / 4.6 / 1232 &  18 / 7.1 / 432 \\
 & Dense 2   &  256 / - / 6.6e4 &  131 / 2.0 / 1.8e4 &  104 / 2.5 / 1.2e4 &  56 / 4.6 / 3136 &  27 / 9.5 / 972 \\
 & ACC       &             0.93 &               0.94 &               0.94 &             0.94 &            0.91 \\
 & TP / TPI  &            - / 8 &             16 / 3 &            21 / 2  &           19 / 1 &          21 / 0 \\
 & HEO / MPI &       8.5e4 / 24 &          2.6e4 / 7 &          1.7e4 / 5 &         5226 / 2 &        1857 / 2 \\
\hline
\multirow{6}{*}{\rotatebox{90}{MNIST-AE1}} & Sparsity  &               - &              0.49 &              0.58 &             0.77 &             0.89 \\
 & Dense 0   &  32 / - / 5.0e4 &  16 / 2.0 / 2.5e4 &  13 / 2.5 / 2.0e4 &   7 / 4.6 / 1.1e4 &  3 / 10.7 / 4704 \\
 & MSE       &            0.02 &              0.03 &              0.03 &              0.04 &             0.05 \\
 & TP / TPI  &           - / 5 &            21 / 3 &            21 / 3 &            21 / 2 &           22 / 2 \\
 & HEO / MPI &      1.0e5 / 14 &         5.0e4 / 7 &         4.1e4 / 6 &         2.2e4 / 5 &         9411 / 4 \\
\hline
\multirow{6}{*}{\rotatebox{90}{MNIST-AE2}} & Sparsity  &               - &              0.47 &              0.57 &             0.78 &             0.88 \\
 & Dense 0   &  64 / - / 1.0e5 &  34 / 1.9 / 5.3e4 &  27 / 2.4 / 4.2e4 &    14 / 4.6 / 2.2e4 &  7 / 9.1 / 1.1e4 \\
 & MSE       &            0.01 &              0.02 &              0.02 &                0.03 &             0.04 \\
 & TP / TPI  &           - / 7 &            39 / 5 &           39 /  4 &              39 / 3 &           38 / 2 \\
 & HEO / MPI &      2.0e5 / 24 &        1.1e5 / 13 &        8.5e4 / 10 &           4.4e4 / 6 &        2.2e4 / 4 \\
\hline
\multirow{8}{*}{\rotatebox{90}{MNIST-AE3}} & Sparsity  &               - &              0.46 &              0.59 &             0.80 &             0.89 \\
 & Dense 0   &  64 / - / 1.0e5 &  33 / 1.9 / 5.2e4 &  26 / 2.5 / 4.1e4 &  13 / 4.9 / 2.0e4 &  7 / 9.1 / 1.1e4 \\
 & Dense 1   &   32 / - / 4096 &   19 / 1.7 / 1254 &    13 / 2.5 / 676 &     8 / 4.0 / 208 &    3 / 10.7 / 42 \\
 & Dense 2   &   64 / - / 4096 &   37 / 1.7 / 1406 &    27 / 2.4 / 702 &    13 / 4.9 / 208 &     7 / 9.1 / 42 \\
 & MSE       &            0.02 &              0.02 &              0.03 &              0.03 &             0.05 \\
 & TP /TPI       &          - / 40 &           51 / 23 &           50 / 18 &           49 / 14 &          49 / 12 \\
 & HEO / MPI &      2.1e5 / 76 &        1.1e5 / 41 &        8.5e4 / 34 &        4.1e4 / 20 &       2.2e4 / 13 \\
\bottomrule
\end{tabular}

\begin{tablenotes}{}
\item - indicates the value is not applicable for the model; N/A indicates the 
value is not available; Conv: Convolution 
\end{tablenotes}

\end{table*}

%% file: MOFHEI.bbl
\begin{thebibliography}{10}
\providecommand{\url}[1]{#1}
\csname url@samestyle\endcsname
\providecommand{\newblock}{\relax}
\providecommand{\bibinfo}[2]{#2}
\providecommand{\BIBentrySTDinterwordspacing}{\spaceskip=0pt\relax}
\providecommand{\BIBentryALTinterwordstretchfactor}{4}
\providecommand{\BIBentryALTinterwordspacing}{\spaceskip=\fontdimen2\font plus
\BIBentryALTinterwordstretchfactor\fontdimen3\font minus
  \fontdimen4\font\relax}
\providecommand{\BIBforeignlanguage}[2]{{%
\expandafter\ifx\csname l@#1\endcsname\relax
\typeout{** WARNING: IEEEtranS.bst: No hyphenation pattern has been}%
\typeout{** loaded for the language `#1'. Using the pattern for}%
\typeout{** the default language instead.}%
\else
\language=\csname l@#1\endcsname
\fi
#2}}
\providecommand{\BIBdecl}{\relax}
\BIBdecl

\bibitem{gdpr}
\BIBentryALTinterwordspacing
E.~G. D. P.~R. 2016. (2016) Regulation (eu) 2016/679 of the european parliament
  and of the council of 27 april 2016 on the protection of natural persons with
  regard to the processing of personal data and on the free movement of such
  data, and repealing directive 95/46/ec (general data protection regulation).
  [Online]. Available: \url{http://data.europa.eu/eli/reg/2016/679/oj}
\BIBentrySTDinterwordspacing

\bibitem{aharoni2011helayers}
E.~Aharoni, A.~Adir, M.~Baruch, N.~Drucker, G.~Ezov, A.~Farkash, L.~Greenberg,
  R.~Masalha, G.~Moshkowich, D.~Murik \emph{et~al.}, ``Helayers: A tile tensors
  framework for large neural networks on encrypted data,'' 2011.

\bibitem{aharoni2023helayers}
------, ``Helayers: A tile tensors framework for large neural networks on
  encrypted data,'' \emph{Proceedings on privacy enhancing technologies}, 2023.

\bibitem{aharoni2022he}
E.~Aharoni, M.~Baruch, P.~Bose, A.~Buyuktosunoglu, N.~Drucker, S.~Pal,
  T.~Pelleg, K.~Sarpatwar, H.~Shaul, O.~Soceanu \emph{et~al.}, ``He-pex:
  Efficient machine learning under homomorphic encryption using pruning,
  permutation and expansion,'' \emph{arXiv preprint arXiv:2207.03384}, 2022.

\bibitem{aharoni2022complex}
E.~Aharoni, N.~Drucker, G.~Ezov, H.~Shaul, and O.~Soceanu, ``Complex encoded
  tile tensors: Accelerating encrypted analytics,'' \emph{IEEE Security \&
  Privacy}, vol.~20, no.~5, pp. 35--43, 2022.

\bibitem{akleylek_efficient_2016}
S.~Akleylek, N.~Bindel, J.~Buchmann, J.~Krämer, and G.~A. Marson, ``An
  efficient lattice-based signature scheme with provably secure
  instantiation,'' in \emph{International {Conference} on {Cryptology} in
  {Africa}}.\hskip 1em plus 0.5em minus 0.4em\relax Springer, 2016, pp. 44--60.

\bibitem{armknecht_guide_2015}
F.~Armknecht, C.~Boyd, C.~Carr, A.~Jaschke, and C.~A. Reuter,
  ``\BIBforeignlanguage{en}{A {Guide} to {Fully} {Homomorphic} {Encryption}},''
  \emph{\BIBforeignlanguage{en}{Cryptology ePrint Archive}}, p.~35, 2015.

\bibitem{arzamasov2018towards}
V.~Arzamasov, K.~B{\"o}hm, and P.~Jochem, ``Towards concise models of grid
  stability,'' in \emph{2018 IEEE International Conference on Communications,
  Control, and Computing Technologies for Smart Grids (SmartGridComm)}.\hskip
  1em plus 0.5em minus 0.4em\relax IEEE, 2018, pp. 1--6.

\bibitem{bellare2012foundations}
M.~Bellare, V.~T. Hoang, and P.~Rogaway, ``Foundations of garbled circuits,''
  in \emph{Proceedings of the 2012 ACM conference on Computer and
  Communications security}, 2012, pp. 784--796.

\bibitem{blalock2020state}
D.~Blalock, J.~J. Gonzalez~Ortiz, J.~Frankle, and J.~Guttag, ``What is the
  state of neural network pruning?'' \emph{Proceedings of machine learning and
  systems}, vol.~2, pp. 129--146, 2020.

\bibitem{brakerski2014leveled}
Z.~Brakerski, C.~Gentry, and V.~Vaikuntanathan, ``(leveled) fully homomorphic
  encryption without bootstrapping,'' \emph{ACM Transactions on Computation
  Theory (TOCT)}, vol.~6, no.~3, pp. 1--36, 2014.

\bibitem{brakerski_leveled_2014}
------, ``({Leveled}) fully homomorphic encryption without bootstrapping,''
  \emph{ACM Transactions on Computation Theory (TOCT)}, vol.~6, no.~3, pp.
  1--36, 2014, publisher: ACM New York, NY, USA.

\bibitem{brassard1987all}
G.~Brassard, C.~Cr{\'e}peau, and J.-M. Robert, ``All-or-nothing disclosure of
  secrets,'' in \emph{Advances in Cryptology—CRYPTO’86: Proceedings
  6}.\hskip 1em plus 0.5em minus 0.4em\relax Springer, 1987, pp. 234--238.

\bibitem{brutzkus2019low}
A.~Brutzkus, R.~Gilad-Bachrach, and O.~Elisha, ``Low latency privacy preserving
  inference,'' in \emph{International Conference on Machine Learning}.\hskip
  1em plus 0.5em minus 0.4em\relax PMLR, 2019, pp. 812--821.

\bibitem{cai2022hunter}
Y.~Cai, Q.~Zhang, R.~Ning, C.~Xin, and H.~Wu, ``Hunter: He-friendly structured
  pruning for efficient privacy-preserving deep learning,'' in
  \emph{Proceedings of the 2022 ACM on Asia Conference on Computer and
  Communications Security}, 2022, pp. 931--945.

\bibitem{cheon_full_2018}
J.~H. Cheon, K.~Han, A.~Kim, M.~Kim, and Y.~Song, ``A full {RNS} variant of
  approximate homomorphic encryption,'' in \emph{International {Conference} on
  {Selected} {Areas} in {Cryptography}}.\hskip 1em plus 0.5em minus 0.4em\relax
  Springer, 2018, pp. 347--368.

\bibitem{cheon_homomorphic_2017}
J.~H. Cheon, A.~Kim, M.~Kim, and Y.~Song, ``Homomorphic encryption for
  arithmetic of approximate numbers,'' in \emph{International {Conference} on
  the {Theory} and {Application} of {Cryptology} and {Information}
  {Security}}.\hskip 1em plus 0.5em minus 0.4em\relax Springer, 2017, pp.
  409--437.

\bibitem{chillotti2020tfhe}
I.~Chillotti, N.~Gama, M.~Georgieva, and M.~Izabach{\`e}ne, ``Tfhe: fast fully
  homomorphic encryption over the torus,'' \emph{Journal of Cryptology},
  vol.~33, no.~1, pp. 34--91, 2020.

\bibitem{chou_faster_2018}
E.~Chou, J.~Beal, D.~Levy, S.~Yeung, A.~Haque, and L.~Fei-Fei, ``Faster
  cryptonets: {Leveraging} sparsity for real-world encrypted inference,''
  \emph{arXiv preprint arXiv:1811.09953}, 2018.

\bibitem{dowlin2016cryptonets}
N.~Dowlin, R.~Gilad-Bachrach, K.~Laine, K.~Lauter, and M.~Naehrig,
  ``Cryptonets: Applying neural networks to encrypted data with high throughput
  and accuracy,'' in \emph{International Conference on Machine Learning}.\hskip
  1em plus 0.5em minus 0.4em\relax PMLR, 2016, pp. 201--210.

\bibitem{fan_somewhat_2012}
J.~Fan and F.~Vercauteren, ``Somewhat {Practical} {Fully} {Homomorphic}
  {Encryption}.'' \emph{IACR Cryptol. ePrint Arch.}, vol. 2012, p. 144, 2012,
  publisher: Citeseer.

\bibitem{frankle2018lottery}
J.~Frankle and M.~Carbin, ``The lottery ticket hypothesis: Finding sparse,
  trainable neural networks,'' \emph{arXiv preprint arXiv:1803.03635}, 2018.

\bibitem{gentry_fully_2009}
\BIBentryALTinterwordspacing
C.~Gentry, ``\BIBforeignlanguage{en}{Fully homomorphic encryption using ideal
  lattices},'' in \emph{\BIBforeignlanguage{en}{Proceedings of the 41st annual
  {ACM} {Symposium} on theory of computing - {STOC} '09}}.\hskip 1em plus 0.5em
  minus 0.4em\relax Bethesda, MD, USA: ACM Press, 2009, p. 169. [Online].
  Available: \url{http://portal.acm.org/citation.cfm?doid=1536414.1536440}
\BIBentrySTDinterwordspacing

\bibitem{ghodsi2020cryptonas}
Z.~Ghodsi, A.~K. Veldanda, B.~Reagen, and S.~Garg, ``Cryptonas: Private
  inference on a relu budget,'' \emph{Advances in Neural Information Processing
  Systems}, vol.~33, pp. 16\,961--16\,971, 2020.

\bibitem{guo2016dynamic}
Y.~Guo, A.~Yao, and Y.~Chen, ``Dynamic network surgery for efficient dnns,''
  \emph{Advances in neural information processing systems}, vol.~29, 2016.

\bibitem{han2015deep}
S.~Han, H.~Mao, and W.~J. Dally, ``Deep compression: Compressing deep neural
  networks with pruning, trained quantization and huffman coding,'' \emph{arXiv
  preprint arXiv:1510.00149}, 2015.

\bibitem{han2015learning}
S.~Han, J.~Pool, J.~Tran, and W.~Dally, ``Learning both weights and connections
  for efficient neural network,'' \emph{Advances in neural information
  processing systems}, vol.~28, 2015.

\bibitem{hesamifard2017cryptodl}
E.~Hesamifard, H.~Takabi, and M.~Ghasemi, ``Cryptodl: Deep neural networks over
  encrypted data,'' \emph{arXiv preprint arXiv:1711.05189}, 2017.

\bibitem{cheetah}
\BIBentryALTinterwordspacing
Z.~Huang, W.~jie Lu, C.~Hong, and J.~Ding, ``Cheetah: Lean and fast secure
  {Two-Party} deep neural network inference,'' in \emph{31st USENIX Security
  Symposium (USENIX Security 22)}.\hskip 1em plus 0.5em minus 0.4em\relax
  Boston, MA: USENIX Association, Aug. 2022, pp. 809--826. [Online]. Available:
  \url{https://www.usenix.org/conference/usenixsecurity22/presentation/huang-zhicong}
\BIBentrySTDinterwordspacing

\bibitem{jacob2018quantization}
B.~Jacob, S.~Kligys, B.~Chen, M.~Zhu, M.~Tang, A.~Howard, H.~Adam, and
  D.~Kalenichenko, ``Quantization and training of neural networks for efficient
  integer-arithmetic-only inference,'' in \emph{Proceedings of the IEEE
  conference on computer vision and pattern recognition}, 2018, pp. 2704--2713.

\bibitem{joseph2020programmable}
V.~Joseph, G.~L. Gopalakrishnan, S.~Muralidharan, M.~Garland, and A.~Garg, ``A
  programmable approach to neural network compression,'' \emph{IEEE Micro},
  vol.~40, no.~5, pp. 17--25, 2020.

\bibitem{juvekar2018gazelle}
C.~Juvekar, V.~Vaikuntanathan, and A.~Chandrakasan, ``$\{$GAZELLE$\}$: A low
  latency framework for secure neural network inference,'' in \emph{27th
  $\{$USENIX$\}$ Security Symposium ($\{$USENIX$\}$ Security 18)}, 2018, pp.
  1651--1669.

\bibitem{krizhevsky2009learning}
A.~Krizhevsky and G.~Hinton, ``Learning multiple layers of features from tiny
  images,'' in \emph{Proceedings of the 25th international conference on
  Machine learning}, 2009, pp. 1097--1104.

\bibitem{lecun1998gradient}
Y.~LeCun, L.~Bottou, Y.~Bengio, and P.~Haffner, ``Gradient-based learning
  applied to document recognition,'' \emph{Proceedings of the IEEE}, vol.~86,
  no.~11, pp. 2278--2324, 1998.

\bibitem{lecun-mnisthandwrittendigit-2010}
\BIBentryALTinterwordspacing
Y.~LeCun, C.~Cortes, and C.~Burges, ``Mnist handwritten digit database,''
  \emph{ATT Labs}, 2010. [Online]. Available:
  \url{http://yann.lecun.com/exdb/mnist/}
\BIBentrySTDinterwordspacing

\bibitem{lee2022privacy}
J.-W. Lee, H.~Kang, Y.~Lee, W.~Choi, J.~Eom, M.~Deryabin, E.~Lee, J.~Lee,
  D.~Yoo, Y.-S. Kim \emph{et~al.}, ``Privacy-preserving machine learning with
  fully homomorphic encryption for deep neural network,'' \emph{IEEE Access},
  vol.~10, pp. 30\,039--30\,054, 2022.

\bibitem{lehmkuhl2021muse}
R.~Lehmkuhl, P.~Mishra, A.~Srinivasan, and R.~A. Popa, ``Muse: Secure inference
  resilient to malicious clients.'' in \emph{USENIX Security Symposium}, 2021,
  pp. 2201--2218.

\bibitem{liu2017learning}
Z.~Liu, J.~Li, Z.~Shen, G.~Huang, S.~Yan, and C.~Zhang, ``Learning efficient
  convolutional networks through network slimming,'' in \emph{Proceedings of
  the IEEE international conference on computer vision}, 2017, pp. 2736--2744.

\bibitem{lou2021hemet}
Q.~Lou and L.~Jiang, ``Hemet: a homomorphic-encryption-friendly
  privacy-preserving mobile neural network architecture,'' in
  \emph{International conference on machine learning}.\hskip 1em plus 0.5em
  minus 0.4em\relax PMLR, 2021, pp. 7102--7110.

\bibitem{mcmahan2017communication}
H.~B. McMahan, E.~Moore, D.~Ramage, S.~Hampson, and B.~Ag{\"u}era~y Arcas,
  ``Communication-efficient learning of deep networks from decentralized
  data,'' in \emph{Proceedings of the 20th International Conference on
  Artificial Intelligence and Statistics (AISTATS)}.\hskip 1em plus 0.5em minus
  0.4em\relax PMLR, 2017, pp. 1273--1282.

\bibitem{neyshabur2017exploring}
B.~Neyshabur, S.~Bhojanapalli, D.~McAllester, and N.~Srebro, ``Exploring
  generalization in deep learning,'' \emph{Advances in neural information
  processing systems}, vol.~30, 2017.

\bibitem{hipaa}
\BIBentryALTinterwordspacing
T.~U.~D. of~Health and H.~S. (HHS). (1996) The health insurance portability and
  accountability act of 1996 (hipaa). [Online]. Available:
  \url{https://www.cdc.gov/phlp/publications/topic/hipaa.html}
\BIBentrySTDinterwordspacing

\bibitem{tfmot-sparsity-prune-low-magnitude}
T.~M. Optimization, ``Prune low magnitude,''
  \url{https://www.tensorflow.org/model_optimization/api_docs/python/tfmot},
  2021, accessed on February 16, 2023.

\bibitem{podschwadt2024memory}
R.~Podschwadt, P.~Ghazvinian, M.~GhasemiGol, and D.~Takabi, ``Memory efficient
  privacy-preserving machine learning based on homomorphic encryption,'' in
  \emph{International Conference on Applied Cryptography and Network
  Security}.\hskip 1em plus 0.5em minus 0.4em\relax Springer, 2024, pp.
  313--339.

\bibitem{ran2023spencnn}
R.~Ran, X.~Luo, W.~Wang, T.~Liu, G.~Quan, X.~Xu, C.~Ding, and W.~Wen,
  ``Spencnn: orchestrating encoding and sparsity for fast homomorphically
  encrypted neural network inference,'' in \emph{International Conference on
  Machine Learning}.\hskip 1em plus 0.5em minus 0.4em\relax PMLR, 2023, pp.
  28\,718--28\,728.

\bibitem{rivlin_chebyshev_2020}
T.~J. Rivlin, \emph{Chebyshev polynomials}.\hskip 1em plus 0.5em minus
  0.4em\relax Courier Dover Publications, 2020.

\bibitem{schroff2015facenet}
F.~Schroff, D.~Kalenichenko, and J.~Philbin, ``Facenet: A unified embedding for
  face recognition and clustering,'' in \emph{Proceedings of the IEEE
  conference on computer vision and pattern recognition}, 2015, pp. 815--823.

\bibitem{sealcrypto}
``{M}icrosoft {SEAL} (release 3.7),'' \url{https://github.com/Microsoft/SEAL},
  Sep. 2021, microsoft Research, Redmond, WA.

\bibitem{shamir1979share}
A.~Shamir, ``How to share a secret. commun. acm,'' 1979.

\bibitem{smart2010fully}
N.~P. Smart and F.~Vercauteren, ``Fully homomorphic encryption with relatively
  small key and ciphertext sizes,'' in \emph{International Workshop on Public
  Key Cryptography}.\hskip 1em plus 0.5em minus 0.4em\relax Springer, 2010, pp.
  420--443.

\bibitem{smart_fully_2014}
------, ``Fully homomorphic {SIMD} operations,'' \emph{Designs, codes and
  cryptography}, vol.~71, no.~1, pp. 57--81, 2014, publisher: Springer.

\bibitem{wang2018rafiki}
W.~Wang, S.~Wang, J.~Gao, M.~Zhang, G.~Chen, T.~K. Ng, and B.~C. Ooi, ``Rafiki:
  Machine learning as an analytics service system,'' \emph{arXiv preprint
  arXiv:1804.06087}, 2018.

\bibitem{wang2017chestx}
X.~Wang, Y.~Peng, L.~Lu, Z.~Lu, M.~Bagheri, and R.~Summers, ``Chestx-ray8:
  Hospital-scale chest x-ray database and benchmarks on weakly-supervised
  classification and localization of common thorax diseases,''
  \emph{Proceedings of the IEEE Conference on Computer Vision and Pattern
  Recognition}, pp. 3462--3471, 2017.

\bibitem{wang2023differential}
Y.~Wang, Q.~Wang, L.~Zhao, and C.~Wang, ``Differential privacy in deep
  learning: Privacy and beyond,'' \emph{Future Generation Computer Systems},
  vol. 148, pp. 408--424, 2023.

\bibitem{yao1982protocols}
A.~C. Yao, ``Protocols for secure computations,'' in \emph{Proceedings of the
  23rd Annual Symposium on Foundations of Computer Science (FOCS)}.\hskip 1em
  plus 0.5em minus 0.4em\relax IEEE, 1982, pp. 160--164.

\bibitem{zhu2017prune}
M.~Zhu and S.~Gupta, ``To prune, or not to prune: exploring the efficacy of
  pruning for model compression,'' \emph{arXiv preprint arXiv:1710.01878},
  2017.

\end{thebibliography}
